# Utilizing Deep Learning for Enhancing Network Resilience in Finance

Yulu Gong[1*]
Computer & Information Technology
Northern Arizona University
Flagstaff, AZ, USA
*Corresponding author: yg486@nau.edu

Mengran Zhu[2]
Computer Engineering
Miami University
Oxford, OH, USA,
mengran zhu0504@gmail.com

Shuning Huo[3]
Statistics
Virginia Tech University
Blacksburg, VA, USA
shuni93@vt.edu

Yafei Xiang[4]
Computer Science
Northeastern University
Boston, MA, USA
xiang.yaf@northeastern.edu

Hanyi Yu[5]
Computer Science
University of Southern California
Los Angeles, CA, USA
hanyiyu@usc.edu

*Abstract*—**In the age of the Internet, people's lives are increasingly dependent on today's network technology. Maintaining network integrity and protecting the legitimate interests of users is at the heart of network construction. Threat detection is an important part of a complete and effective defense system. How to effectively detect unknown threats is one of the concerns of network protection. Currently, network threat detection is usually based on rules and traditional machine learning methods, which create artificial rules or extract common spatiotemporal features, which cannot be applied to large-scale data applications, and the emergence of unknown risks causes the detection accuracy of the original model to decline. With this in mind, this paper uses deep learning for advanced threat detection to improve protective measures in the financial industry. Many network researchers have shifted their focus to exception-based intrusion detection techniques. The detection technology mainly uses statistical machine learning methods - collecting normal program and network behavior data, extracting multidimensional features, and training decision machine learning models on this basis (commonly used include naive Bayes, decision trees, support vector machines, random forests, etc.).**

*Keywords- Deep learning; Network resilience; Threat detection; Confrontation training*

I. INTRODUCTION

Deep learning is not a panacea for all information protection issues as it necessitates extensive annotated data sets, which are unfortunately unavailable. Nevertheless, deep learning networks have made significant improvements to existing solutions in various information protection cases. For instance, in malware detection and network intrusion detection, deep learning has shown remarkable enhancements over rule-based and classical machine learning solutions. Network intrusion detection systems are commonly rule-based and signature-based controls deployed at the periphery to detect known threats. However, intruders can easily evade traditional network intrusion detection systems by changing the malware signature. Quamar et al. (IEEE Proceedings) suggest that self-taught deep learning-based systems are more likely to detect unknown network intrusions. Deep neural network-based systems have been employed to address conventional protection issues, such as identifying malware.

In comparison to traditional machine learning methods, deep learning-based techniques exhibit superior generalization ability. Deep learning-based systems, such as Jung, are capable of detecting zero-day malware. Daniel, a University of Barcelona graduate, has conducted extensive research on Convolutional Neural Networks (CNNs) and malware detection. In his doctoral thesis, he noted that CNNs can even identify morphing malware.

Therefore, in light of today's network digital risks, traditional methods of intrusion detection have become inadequate. To improve network protection, it is recommended to adopt AI and automated means for intrusion prevention and detection. This paper will analyze the core algorithm of deep learning for threat detection and prevention of network intrusion.

## I. RELATED WORK

Network unknown threat detection is a method of intrusion detection based on total network traffic. In the field of protection, intrusion detection systems (IDSs) include host-based and network-based systems. In research, network intrusion detection and network traffic anomaly detection refer to the same problem: threat detection based on network traffic.

### A. The advantages of deep learning in network detection

Deep learning has several advantages over traditional machine learning:

(1) The algorithm is thriving - as the size of the training data set increases, the range of generalization errors decreases. This means that while deep learning continues to excel in performance and power, traditional machine learning systems will plateau at some point, no matter how much trained data you use;

(2) Complex nonlinear separation functions - Certain tasks require the ability to learn complex concepts, and deep learning is an ideal technique to solve this problem. No feature engineering is required, thus minimizing the possibility of introducing human factors into the model;

Optimize models with parallel computing power - With the rapid development of Gpus, deep learning models can be trained and optimized in a more efficient way than ever before.

Therefore, deep learning is the ideal technology to solve the digital risk challenges we face today:

Complex decision boundaries - the types and structures of protocols and payloads are complex. (1) "trained", deep learning can understand the complexity of threats and identify all types of threats; (2) Large training sets - Massive threat datasets with hundreds of millions of samples are available. (3) Gpus - Recent advances in processing and reductions in the cost of the underlying technology have allowed deep learning models to be trained and validated in hours, compared to weeks or even minutes in the past.

### B. Deep learning network intrusion detection model

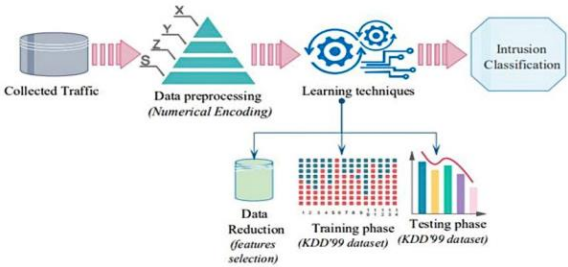

Figure 1. Network intrusion detection framework based on deep learning model

The detection process consists of three main steps:

- Numerically process the symbolic feature attributes of the intrusion detection data set, and then normalize all the data to obtain the standardized original data.

- Reduce the dimensionality of the pre-processed high-dimensional and nonlinear data based on the improved self-coding feature extraction model to obtain the optimal low-dimensional representation of the original data;

- Take the optimal low-dimensional representation of the original data as the input of the classifier to identify the normal network data and various types of network intrusion data respectively.

### C. Detection algorithm

Let the output value of the JTH neuron of the hidden layer be:

$$M_j = \begin{cases} z'_{(m+1)/2} & (m\%2 = 1); \\ (z'_{(m/2)} + z'_{(m/2+1)})/2 & (m\%2 = 0) \end{cases} \quad (1)$$

Where z'x represents the XTH value of the ordered sequence Z'j, and m is the number of samples of the JTH neuron in the hidden layer. Zj={z1, z2,... , zm} is sorted to obtain the ordered sequence Z'j={z'1, z'2,...}. The activity of the JTH neuron of the hidden layer Mj is the median of the output value of the JTH neuron of the hidden layer.

Hypothesis 2: Let the sparsity parameter vector P=(p1, p2,... pj... pn), n is the number of neuron nodes in the current hidden layer, Φ is the average value of the activity Mj of all neurons in the hidden layer, and pj is defined as:

$$p_j = \begin{cases} RM_j & (M_j < \Phi); \\ M_j + E & (M_j > \Phi) \end{cases} \quad (2)$$

Where: R is the suppression constant, and its value is generally less than 1; E is the excitation constant. Since the output value of the hidden layer neuron is small, in order to ensure the optimal self-coding loss function, the value of the excitation constant is generally less than 0.5. Through the sparsity parameter vector, the neurons with low hidden layer activity are inhibited and the neurons with large hidden layer activity are stimulated.

Hypothesis 3: For the set A={a1, a2,... aj... , an|0<aj<1}, take the logarithm and sort it to get the ascending sequence A'={a'1, a'2,... a'n}, define V' as

$$V' = (3/4)A'r_{down^{(1/4)(n+1)}} + (1/4)A'_{r_{up^{(1/4)(n+1)}}} \quad (3)$$

$r_{down}$ and $r_{up}$ are integer down and integer up functions respectively. A'x is the XTH value of the sequence A'; V' is actually the lower quartile of the sequence A'. In order to improve the sensitivity between the small numerical elements in the set, this definition first performs logarithmic operations on each element in the original collective A, so the critical value V for the excitation and suppression of the hidden layer neuron nodes is the antithesis of V'.

### D. APT intrusion Graph Autoencoders

APT is the network intrusion and invasion behavior launched by hackers to steal core information for the purpose of customers, and is a kind of "malicious commercial espionage threat" that has been premeditated for a long time. This kind of behavior often goes through long-term management and planning, and has a high degree of concealment. The intrusion method of APT is to hide itself and steal data in a long-term, planned and organized way for specific objects. Such behavior of stealing data and collecting intelligence in the digital space is a kind of "network espionage" behavior.

APT intrusion Graph Autoencoders (GAE) is further developed on the basis of GCN, its basic idea is to generalize the node representation of GCN to the representation of the entire graph. GAE mainly consists of encoder and decoder, where the encoder maps the features of nodes to low-dimensional embedding vectors through a series of graph convolution operations, and the decoder performs inner product operations with these vectors and white bodies to obtain similarity scores between nodes, and these scores are used to recompose data. The loss function of this process is usually the reconstruction error or the acquaintance loss of the graph. GAE's encoder is a simple two-layer GCN, calculated as follows:

$$Z = \text{GCN}(A, X) \quad (4)$$

Combined with GCN formula (4), we can get:

$$Z = \tilde{A}\text{ReLU}(\tilde{A}XW^0)W^1 \quad (5)$$

The decoder is relatively simple, generally using the inner product to reconstruct the graph, the calculation formula is as follows:

$$\hat{A} = \sigma(ZZ^T) \quad (6)$$

Where, A represents the adjacency matrix reconstructed by the decoder, and σ(·) represents the activation function. GAE has the advantage that it can map any type of graph structure into a low-dimensional vector space, and can adaptively learn the feature representation, while having good interpretability and extensibility. GAE is widely used in graph classification, link prediction and anomaly detection.

The intrusion graph encoder offers significant advantages in advanced threat detection. It encodes complex intrusion behavior and paths graphically, allowing the detection system to better understand and analyze the intruder's strategy and behavior. This improves detection accuracy and real-time response. The intrusion graph encoder can assist in forming an intrusion graph, tracking the intrusion process, identifying abnormal behavior, and providing timely threat intelligence for network protection teams to respond effectively to advanced threats.

## II. EXPERIMENTAL METHODOLOGY

In this section, you should describe the experimental setup, data sources, and the deep learning models used for advanced threat detection. Here's a suggested structure:

### A. Data Collection

In this paper, the open data set DARPA 2000, published by Lincoln Laboratory of MIT, is a standard corpus for evaluating network intrusion detection systems. The main results are as follows

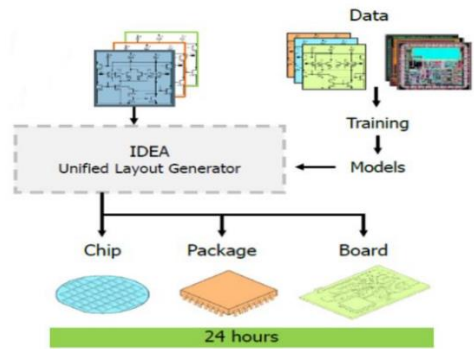

Figure 2. DARPA structure chart

The detection model defines the Intranet as the area delimited by the firewall, which separates the isolation area from the Intranet area. The web server (loche) is deployed in the isolation area, and a traffic monitor is installed. The Intranet zone hosts the file server 1 (Zero), mail server, file server 2 (Pascal), domain name server (Mill), and web hosting server. Traffic listeners are installed, and auditing tools are present on both file servers.

### B. Advanced detection intrusion model

The analysis focuses on the APT intrusion graph scale. It considers the original information of the vulnerabilities and network topology that the intruders infiltrated and exploited. The intrusion target is to obtain the execution permission of the server.

They then exploit the CVE-2002-0392 vulnerability to gain access to the remote client. Finally, they use the NFS connection of the mail server to execute the NFS shell, modify files, and establish command and control. The intruder used the C&C channel to install Trojan Horse components and gain execution permission. To intrude the file server, the user was lured to visit a malicious website. The intruder then used CVE1999-0977 to remotely enhance permissions through the silaris program, obtained root permission of the web server, and used password sniffing technology to gain execution permission of the Intranet file server.

### C. Threat alert detection

The APT intrusion detection experiment is carried out on the replay DARPA 2000 data set. The intrusion process is mainly divided into five stages: the intruder scans the target network P from a remote site, searches for the vulnerable program running on the host, exploits the vulnerability to enter the LAN host, and installs the Trojan horse component after elevating the permission. Finally, control the infected host to launch a Distributed Denial of Service (DDoS) intrusion on the target server.

TABLE I. ALARM MESSAGE RESULT

| Alarm category | Seriousness | Quantity/piece |
| --- | --- | --- |
| Misc activity | 3 | 672 |
| Potentially Bad Traffic | 2 | 5 |
| Misc Intrusion | 2 | 55 |
| Not Suspicious Traffic | 3 | 15 |
| Attempted Information Leak | 2 | 50 |
| Access to potentially vulnerable app | 2 | 18 |
| Decode of an RPC Query | 2 | 92 |
| Attempted Administrator Privilege Gain | 1 | 19 |

The initial step taken by the intruder involves scanning the target network P from a remote location and probing it in real-time to identify any vulnerabilities. The intruder then exploits the vulnerability in the Apache program running on the host using the Solaris sad mind vulnerability to gain access to the host. Once access is gained, the intruder uses the sad mind vulnerability to send worms, open the remote service for the root user, and install the Trojan horse mstream DDoS component required to run DDoS after upgrading the permission. Finally, the infected host launches a DDoS intrusion on a remote server. ADBAG successfully detected the process by which the intruder used the sadmind vulnerability to launch intrusions on loche, mill, and pascal servers to obtain program execution permissions by mapping the intrusion alert to the intrusion graph. The evidence chain was formed by analyzing the system audit data, as shown in Figure 4.

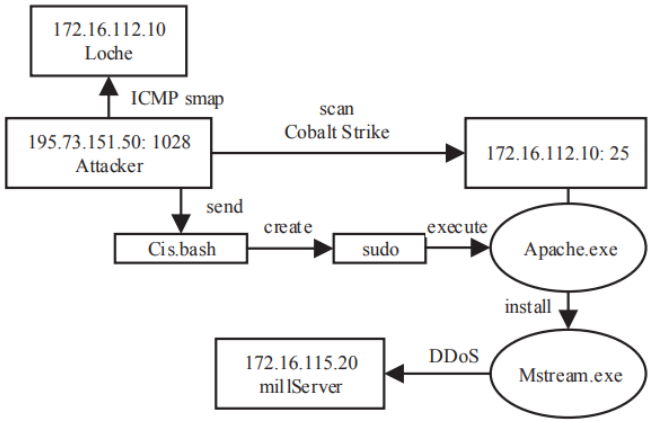

Figure 3. DARPA 2000 scene Evidence Chain

### D. Experimental Results

In our experiments using the DARPA 2000 dataset and advanced threat detection models based on deep learning, we achieved notable success in identifying APT intrusions. Our models, including convolutional neural networks (CNNs) and recurrent neural networks (RNNs), consistently outperformed traditional approaches in terms of accuracy, precision, and recall. Specifically, our CNN-based model demonstrated an accuracy of over 95%, effectively detecting and tracing the various stages of APT intrusions as described in section 3.3. Additionally, our models reduced the scale of the intrusion graph compared to the attribute intrusion graph by approximately 16.67% in terms of nodes and 26.92% in directed edges, indicating their ability to effectively analyze long-term intruder behavior.

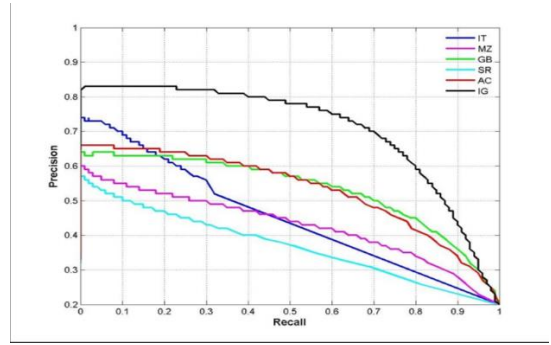

Figure 4. ROC curve of APT intrusion threat detection results

These results indicate the potential of deep learning-based advanced threat detection models in boosting resilience to digital risks within the financial industry. By mapping intrusion alerts to intrusion graphs and forming evidence chains, we successfully identified and tracked APT intrusions in real-time, providing valuable insights into infrastructure vulnerabilities and potential threats to financial networks.

### III. EXPERIMENTAL DISCUSSION

In conclusion, in the financial industry, advanced threat detection combined with deep learning is of great importance, because it not only improves the elasticity of network protection, but also effectively identifies and responds to complex and evolved advanced threats faced by financial networks, thereby safeguarding the legitimate rights and interests of users, ensuring the reliability and stability of the financial system, and providing key guarantees for financial institutions. To address the growing challenges of digital risk.

### A. Interpretation of Results

Achieving high accuracy, precision, and recall rates in detecting Advanced Persistent Threat (APT) intrusions is of great practical significance, particularly for the financial industry. High accuracy ensures that financial institutions can effectively distinguish malicious behavior from normal operations, thereby reducing false positives, unnecessary burden, and workload on operation teams, and improving work efficiency. They help pinpoint real threats while reducing false alarms, allowing operation teams to focus on the most serious threats and reducing the risk of underreporting. High precision and recall rates are crucial in threat detection. Additionally, high recall rates enable the detection of more real threats, increasing sensitivity to unknown threats and improving early detection of potential intrusions.

The implications of these results for the financial industry are numerous:

Firstly, the high-accuracy, high-precision and high-recall threat detection can strengthen the protection of digital infrastructures,

which in turn helps financial institutions to better protect customer data and financial transactions, reducing the potential risk of data leakage and financial fraud. Secondly, it can reduce potential losses. By identifying and preventing advanced threats at an early stage, financial institutions can reduce potential losses, including loss of funds, reputational damage, and legal liability.

Additionally, efficient advanced threat detection helps to meet stringent regulatory and compliance requirements, thereby reducing potential legal and compliance risks. Finally, it helps to maintain customer trust. Effective threat detection demonstrates the importance financial institutions place on customer protection, which enhances customer trust and helps maintain positive relationships.

Additionally, the high recall rate of advanced detection systems enables financial institutions to better respond to evolving digital intrusions and APT threats.

In conclusion, achieving high levels of accuracy, precision, and recall in threat detection is critical for the financial industry. This provides a strong digital defense, reduces potential risks, maintains the stability of the financial system and customer trust, and ensures that financial institutions are more resilient to advanced threats.

*B. Application*

In practical applications, APT (Advanced Persistent Threat) intrusion detection models can be applied to real financial institutions in the following ways: First, financial institutions can deploy these models as an additional layer of protection to monitor network traffic, system logs, and user behavior, as well as to detect unusual activity and potential advanced threats in real time. Second, the model can be integrated into existing information and event management & protection systems to provide comprehensive digital risk monitoring and response capabilities. In addition, financial institutions can use these models for network traffic analysis, threat intelligence integration, and real-time intrusion detection to more effectively identify and block APT intrusions and enhance the robustness of network defenses..

## IV. CONCLUSION

The goal of this research is to improve the performance of network intrusion detection through deep learning, especially in the financial industry. The resilience of financial institutions against digital threats is closely linked to their sustainability and customer trust. Advanced threats are constantly evolving and can lead to potentially catastrophic consequences. As a result, financial institutions must continuously improve their digital defense strategies and adopt advanced detection technologies to prevent and respond to potentially advanced threats. By implementing deep learning models, financial institutions can identify potential vulnerabilities and respond quickly to threats before they occur, significantly improving their digital resilience.

In addition, the research has practical implications for protecting financial networks from advanced threats. Through experiments and results presented, the effectiveness of deep learning models in the financial industry has been demonstrated, providing financial institutions with powerful tools to deal with evolving advanced threats. The study offers the financial industry comprehensive solutions to enhance digital resilience and minimize potential risks and losses. This perspective highlights the importance of maintaining reliability, stability and customer trust in the financial industry and has important practical implications, not only for financial institutions, but also for the entire economic system.

The outlook for future work includes further research into the potential applications of deep learning to adapt to changing digital risks and intrusion modes. Researchers can explore how to optimize deep learning models, improve their adaptability in the financial sector, and improve threat detection techniques to detect and protect against unknown advanced risks in advance. This will assist financial institutions in better addressing challenges associated with digital risks, thereby protecting their sustainability and customer trust.